\documentclass[reprint, amsmath,amssymb,pra,floatfix]{revtex4-2}

\usepackage{graphicx}
\usepackage{dcolumn}
\usepackage{bm}
\usepackage{hyperref}
\usepackage{url}

\usepackage{xcolor}

\newcommand{\red}{\textcolor{red}}

\newcommand{\fracc}[2]{\frac{\textstyle{#1}}{\textstyle{#2}}}

\begin{document}

\preprint{APS/123-QED}

\title{Light propagation in (2+1)-dimensional electrodynamics: the case of nonlinear constitutive laws}%
\author{Eduardo Bittencourt}
 \email{bittencourt@unifei.edu.br}
\author{Elliton O. S. Brandão}%
 \email{ellitonbrandao@unifei.edu.br}
\affiliation{Federal University of Itajub\' a, BPS Avenue, 1303, Itajub\'a/MG - Brazil
}
\author{\'Erico Goulart}
 \email{egoulart@ufsj.edu.br}
\affiliation{Federal University of S\~ao Jo\~ao d'El Rei, C.A.P. Rod.: MG 443, KM 7, Ouro Branco/MG - Brazil
}

\date{\today}

\begin{abstract}
We scrutinize the geometrical properties of light propagation inside a nonlinear medium modeled by a fully covariant electromagnetic theory in $2+1$-dimensions. After setting the nonlinear constitutive relations, the phase velocity and the polarization of waves are derived and three special cases are analyzed in details. In spite of the dimensional reduction, our model still presents phenomena like one-way propagation, controlled opacity among others for a large class of dielectric and magneto-electric parameters.
\end{abstract}

\maketitle


\section{\label{sec:level1} INTRODUCTION}

The last years have witnessed a great interest in the understanding of light propagation inside two-dimensional (2D) materials. The main reason behind this is the unique optical response such mono-layer structured media (of atomic thickness) present under excitation by external electromagnetic fields. This peculiar feature was already noticed from the first studies with graphene \cite{Novoselov2004} and thenceforth, many other 2D materials have been discovered, generating a rapidly growing development of optoeletronic and photonics for these media. Examples of exciting applications include THz wave generation \cite{zhao2017}, passive optical power limiters \cite{varma2017,Feng2010}, electro-optic modulators \cite{Seyler2015,Klein2017} among others (see Ref.\ \cite{You2019} and references therein for a complete list of technological developments).


Nowadays, there is a reasonably large family of known materials presenting a relatively ultrafast and strong optical nonlinearity response like perovskites \cite{Richter2017}, gallium selenide \cite{Karvonen2015}, black phosphorus \cite{Wang2016}, transition metal chalcogenides \cite{Wang2015} and others. The success of these media in paving the way for new optical devices is essentially due to their large optical and thermal damage threshold, with ultrafast recovery time and high chemical/mechanical stability. They also range the whole spectrum of electric conductivity, being (semi) metals, semiconductors or insulators. At last, they are of low fabrication costs.


Inspired by the aforementioned technological achievements, we take a step back and theoretically scrutinize some aspects of light propagation inside nonlinear 2D materials in the limit of geometric optics. Based mostly upon the results derived in Ref.\ \cite{bit_bran_goul_22}, where the effective metric associated to a (2+1)-dimensional electrodynamics with a linear constitutive law was derived, we now turn our attention to nonlinear constitutive laws. Due to the dimension reduction with respect to the original $(3+1)$ Maxwell's theory, it is possible to study the full nonlinear theory following a quite similar approach. More specifically, we determine the general expressions for the phase velocities and the corresponding wave-polarization vectors for nonlinear dielectrics, nonlinear magnetoelectric media and a combination thereof.


This paper is organized as follows. Section II starts with a discussion of the equations of motion and the corresponding nonlinear constitutive laws. Then, we construct the eigenvalue equation associated with the light propagation in a general nonlinear 2D material. In Section III, the expression for the phase velocity is analyzed, showing the possibility of one-directional propagation for sifficiently high values of magneto-electric cross terms. In Section IV, we characterize the eigenvectors of the original eigenvalue equation and derive the corresponding wave-polarization vectors. Finally, in Section V, we study the behavior of the phase speed and the wave-polarization vector for some special nonlinear systems: purely magnetic media, anisotropic non-magnetic dielectrics and second order magneto-electric materials. 
Throughout, we work with a Minkowski background metric expressed in a cartesian coordinate system i.e., $\eta_{ab}=\mbox{diag}(-1, +1, +1)$. As usual, the quantities  $\varepsilon_{0}$ and $\mu_{0}$ respectively denote the electric permittivity and magnetic permeability of the vacuum and  units are chosen
such that $c=1$, except where mentioned otherwise.


\section{\label{sec:level2}Mathematical setting}

Here, we shall follow the steps of our previous work \cite{bit_bran_goul_22}, dealing with the electromagnetic field being represented by a 2-form field $F_{ab}$. The main difference with respect to our previous analysis is the assumption of a timelike observer field that will decompose $F_{ab}$, from the very beginning, in terms of a vector electric field and a (pseudo) scalar magnetic field. Consequently, all the results derived thereafter are observer-dependent, although they are fully covariant. This is very convenient for the analysis of the special cases we shall handle and for further comparison with experiments.

\subsection{Field equations}


We start by considering a flat (2+1)-dimensional spacetime and writing the Faraday tensor in cartesian coordinates $x^{a}=(t,x,y)$ as
\begin{equation}
\label{faraday_def}
F_{ab} = \begin{pmatrix}
0 & -E_x & -E_y\\
E_x & 0 & B \\
E_y & -B & 0
\end{pmatrix}.
\end{equation}
More generally, with the introduction of a future-directed, timelike, normalized field of inertial observers, henceforth denoted by $t^{a}$, this tensor can be irreducibly decomposed as 
\begin{equation}
\label{decomp_far_ten}    
F_{ab}=t_{a}E_{b}-t_{b}E_{a}+B\epsilon_{abc}t^{c},
\end{equation}
where $\epsilon_{abc}$ is the completely skew-symmetric Levi-Civita tensor with $\epsilon_{012}=+1$. As usual, writing the Hodge dual as 
\begin{equation}
\label{decomp_dual_far_ten}
^*F^{a}=\frac{1}{2} \epsilon^{a}{}_{bc}F^{bc}=\epsilon^{a}{}_{bc}t^{b}E^{c}-Bt^{a},
\end{equation}
a direct calculation gives the projections
\begin{equation}
E_{a} = F_{ab}t^{b},\quad\quad\quad  B= ^*F_{c}t^{c},
\end{equation}
which describe, respectively, the (vector) electric field and the (pseudo-scalar) magnetic induction as measured by the corresponding inertial observer. In addition, we write the squared norms of these quantities as
\begin{equation}
||E||^{2}=E^{a}\,E_{a},\quad\quad\quad ||B||^{2}=B^{2},
\end{equation}
and notice that the Faraday tensor and its dual do not share the same tensor rank, which is expected in any spacetime dimension different from $3+1$. 

Inside a nonlinear material medium, we define the ``polarization'' tensor $P_{ab}$ to account for the response of the medium to external applied fields. This tensor is also irreducibly decomposed as
\begin{equation}
\label{decomp_pol_ten}
P_{ab} = t_{a}D_{b} - t_{b}D_{a} + H \epsilon_{abc}t^{c},
\end{equation}
where the field excitations are the (vector) electric displacement $D_{a}$ and the (pseudo-scalar) magnetic field $H$.
It is generally assumed that these field excitations and the field strengths are linked through constitutive relations of the form
\begin{eqnarray}
&&D^{a} = \mathfrak{A}^{a}{}_{b}\,E^{b} + \mathfrak{B}^{a}\,B, \label{const_relat1}\\[1ex]
&&H = \mathfrak{C}_{a} \, E^{a} + \mathfrak{D} \, B, \label{const_relat2}
\end{eqnarray}
with $\mathfrak{A}^{a}{}_{b}$ denoting the electric permittivity tensor, $\mathfrak{D}$ the (inverse) magnetic permeability, and the other terms properly introduced in order to take into account possible magneto-electric effects, to wit, the ``magnetic permittivity'' represented by $\mathfrak{B}^{a}$ and the ``electric permeability'' represented by $\mathfrak{C}_{a}$. The above quantities are spacelike and orthogonal to the observer field by construction and are allowed to depend smoothly on spacetime position as well as on the external electromagnetic fields $\{E^{a}(t,x,y),B(t,x,y)\}$.

The equations of motion inside a (2+1)-dimensional nonlinear medium are defined as 
\begin{equation}
\label{max_eqs}
P^{ab}{}_{,b} = J^{a}\quad \mbox{and} \quad ^{*}F^{a}{}_{,a} = 0,
\end{equation}
where ``\,$,$\,'' means partial derivative and $J^{a}$ is the 3-vector current density. Projecting the latter along the observer field and onto the two-dimensional rest space orthogonal to it, through the projector $h^{a}{}_{b}=\delta^{a}_{b}+t^{a}t_{b}$, yield
\begin{eqnarray}
&&D^{a}{}_{,a}=\rho, \label{eq:1}\\[1ex]
&&h^{a}{}_{b}\dot D^{b} - \epsilon^{ab}{}_{c}\,t^{c}\,H_{,b}=h^{a}{}_{b} J^{b}, \label{eq:2}\\[1ex]
&&\dot B+\epsilon^{a}{}_{bc}\,t^{c}\,E^{b}{}_{,a} = 0,\label{eq:3}
\end{eqnarray}
with $\dot X^{a}\equiv X^{a}{}_{,b}t^{b}$ and $\rho=-J_{a}t^{a}$. Several interesting features of these equations and their main distinctions from the 3+1 case can be found in Refs.\ \cite{Lapidus1982,McDonald2019,DBoito2020,Maggi2022}. From now on, we shall focus on the limit of geometric optics.

The partial derivatives of the constitutive relations\ (\ref{const_relat1}) and (\ref{const_relat2}) are easily calculated as
\begin{eqnarray}
 D^{a}{}_{,b}&=&\tilde{\mathfrak{A}}^{a}{}_{c}\, E^{c}{}_{,b} + \tilde{\mathfrak{B}}^{a}\,B_{,b}+\ldots \label{def_D;mu}\\[1ex]
H_{,a}&=&\tilde{\mathfrak{C}}_{b}\,E^{b}{}_{,a}+\tilde{\mathfrak{D}} B_{,a}+\ldots
\label{def_H;mu}\end{eqnarray}
with the ellipsis standing for algebraic contributions whose explicit form is irrelevant for our discussion and
\begin{eqnarray}
\tilde{\mathfrak{A}}^{a}{}_{b} &=&\mathfrak{A}^{a}{}_{b} + \frac{\partial\mathfrak{A}^{a}{}_{c}}{\partial E^{b}} \, E^{c}\,+\frac{\partial\mathfrak{B}^{a}}{\partial E^{b}}\,B,\label{tildeA}\\[1ex] 
\tilde{\mathfrak{B}}^{a}&=&\mathfrak{B}^{a}+\frac{\partial\mathfrak{A}^{a}{}_{b}}{\partial B}\,E^{b}+\frac{\partial\mathfrak{B}^{a}}{\partial B}\,B, \label{tildeB}\\ [1ex]
\tilde{\mathfrak{C}}_{a}&=&\mathfrak{C}_{a}+\frac{\partial\mathfrak{C}_{b}}{\partial E^{a}}\,E^{b}+\frac{\partial\mathfrak{D}}{\partial E^{a}}\,B,\label{tildeC}\\[1ex]
\tilde{\mathfrak{D}}&=&\mathfrak{D} + \frac{\partial\mathfrak{C}_{a}}{\partial B}\,E^{a}+\frac{\partial\mathfrak{D}}{\partial B}\,B\label{tildeD}.
\end{eqnarray}
Henceforth, for the sake of terminology, we shall call the set $\{\tilde{\mathfrak{A}}^{a}{}_{b}, \tilde{\mathfrak{B}}^{a},\tilde{\mathfrak{C}}_{a},\tilde{\mathfrak{D}}\}$ the \textit{constitutive tetrad} of the nonlinear medium. Clearly, each element of the tetrad will also be a smooth function of position as well as field strengths.

\subsection{Field discontinuities}

In order to study the main features of wave propagation, we consider a wavefront surface $\Sigma(x)=const$ and define the normal co-vector as $k_{a} = \partial{\Sigma}/\partial{x^{a}}$. Assuming the field strengths $E^{a}$ and $B$ are continuous through $\Sigma$, but with a possible non-zero step in their first derivatives, according to Hadamard's theorem \cite{Hadamard1903,papapetrou}, we determine such step as
\begin{equation}
[E^{a}{}_{,b}]\big|_{\Sigma} = e^{a} k_{b},\quad \mbox{and}\quad [B_{,a}]\big|_{\Sigma} = b \, k_{a},    
\end{equation}
where $e^{a}$ and $b$ are amplitudes representing the \textit{wave-polarization vector} and the \textit{wave-polarization scalar}, respectively.

In order to proceed, it is convenient to introduce the following projections
\begin{equation}
\omega \equiv k_{a} t^{a},\quad\quad q_{a} \equiv h_{a}{}^{b}k_{b},
\end{equation}
where $\omega$ is the wave frequency and $q_{a}$ is the corresponding co-vector orthogonal to the observer field. By writing the squared norm of the latter as $||q||^{2} = h_{ab}q^{a}\,q^{b}$, we then define the space-like ortho-normal vectors
\begin{equation}
\hat{q}^{a} = q^{a}/||q||,\quad\quad\quad \hat{p}^{a}=\epsilon^{abc}\hat{q}_{b}t_{c},
\end{equation}
and let the wave co-vector be schematically written as 
\begin{equation}
k_{a}=||q||\ (v_{\varphi},\ \hat{q}_{a})    
\end{equation}
where $v_{\varphi}=\omega/||q||$ is the usual phase velocity as measured by the observer. With these conventions, by first applying Hadamard's step conditions to Eq.\ (\ref{eq:3}), and taking into account Eq.\ (\ref{def_H;mu}), gives
\begin{equation}
v_{\varphi}b=\hat{p}_{c}e^{c}
\end{equation}
Similarly, by applying the step conditions to Eqs.\ (\ref{eq:1})-(\ref{eq:2}) with Eqs.\ (\ref{def_D;mu})-(\ref{def_H;mu}), and using the above relation yields
\begin{eqnarray}
&\left(v_{\varphi}\ \tilde{\mathfrak{A}}^{a}{}_{b}+ A^{a}{}_{b}\right)\hat{q}_{a}\,e^{b}=0,\label{eq:Gauss}\\[1ex]
&\left(v_{\varphi}^{2}\ \tilde{\mathfrak{A}}^{a}{}_{b} + v_{\varphi}\,A^{a}{}_{b} -\,\tilde{\mathfrak{D}}\, \hat{p}^{a}\hat{p}_{b}\right)e^{b} = 0, \label{pre_fresnel}
\end{eqnarray}
where we have introduced the auxiliary magneto-electric mixed quantity $A^{a}{}_{b}=\tilde{\mathfrak{B}}^{a}\,\hat{p}_{b}-\tilde{\mathfrak{C}}_{b}\,\hat{p}^{a}$, for conciseness.

A closer inspection of the above relations reveals that Eq.\ (\ref{eq:Gauss}) is actually a consequence of Eq.\ (\ref{pre_fresnel}), due to the orthogonality of $\hat{q}^{a}$ and $\hat{p}^{a}$. Therefore, if we define the \textit{generalized Fresnel matrix} as
\begin{equation}\label{fm}
Z^{a}{}_{b}\equiv\left(v_{\varphi}^{2}\ \tilde{\mathfrak{A}}^{a}{}_{b} + v_{\varphi}\,A^{a}{}_{b} -\,\tilde{\mathfrak{D}}\, \hat{p}^{a}\hat{p}_{b}\right),
\end{equation}
the problem of wave propagation reduces to the eigenvalue/eigenvector problem
\begin{eqnarray}
\label{eq:13}
   Z^{a}{}_{b}{}e^{b}=0\quad\rightarrow\quad \mbox{rank}(Z^{a}{}_{b})\leq 2.
\end{eqnarray}
However, since the observer field automatically belongs to the kernel of the Fresnel matrix, non-trivial solutions for the polarization will exist provided we require that $\mbox{rank}(Z^{a}{}_{b})=1$.

\section{Phase velocities}

A well known result of linear algebra implies that any real $3\times 3$ matrix with rank one satisfies the algebraic relation
\begin{equation}\label{adj}
\frac{1}{2}(Z^{a}{}_{a}Z^{b}{}_{b}-Z^{a}{}_{b}Z^{b}{}_{a})=0,
\end{equation}
which is a direct consequence of the vanishing of the adjoint matrix. A direct calculation combining Eq.\ (\ref{adj}) with Eq.\ (\ref{fm}) shows that the corresponding fourth-order polynomial somehow factorizes, thus yielding a second-order polynomial equation for the phase velocity, as follows
\begin{equation}
\label{ph_vel}
\alpha v_\varphi^2-\beta v_\varphi-\gamma = 0,
\end{equation}
where the coefficients are defined as
\begin{eqnarray}
&&\alpha = \frac{1}{2}(\tilde{\mathfrak{A}}^{a}{}_{a}\tilde{\mathfrak{A}}^{b}{}_{b}-\tilde{\mathfrak{A}}^{a}{}_{b}\tilde{\mathfrak{A}}^{b}{}_{a}),\label{eq:18}\\[1ex]
&&\beta =\tilde{\mathfrak{A}}\,^{a}{}_{b} A^{b}{}_{a} - \tilde{\mathfrak{A}}^{a}{}_{a}A^{b}{}_{b},\label{eq:19}\\[1ex]
&&\gamma = \tilde{\mathfrak{D}}\,\tilde{\mathfrak{A}}_{ab}\,\hat{q}^{a}\,\hat{q}^{b}-\frac{1}{2}(A^{a}{}_{a}A^{b}{}_{b}-A^{a}{}_{b}A^{b}{}_{a}).\label{eq:20}
\end{eqnarray}
It is worth mentioning that Eq. (\ref{ph_vel}) gives rise to a homogeneous quadratic multivariate polynomial at the cotangent bundle. In the case of linear constitutive laws, the precise form of this polynomial was rigorously derived in Ref.\ \cite{bit_bran_goul_22}, using the method of the effective metric. In principle, such an observer-independent and totally covariant approach could be carried out for the nonlinear case as well. However, since all relevant information about the characteristic surfaces as well as ray propagation is contained in Eq.\ (\ref{ph_vel}), we shall concentrate our efforts solely to the study of phase velocities.

Suppose that the constitutive tetrad $\{\tilde{\mathfrak{A}}^{a}{}_{b}, \tilde{\mathfrak{B}}^{a},\tilde{\mathfrak{C}}_{a},\tilde{\mathfrak{D}}\}$ and a direction $\hat{q}^{a}$ are given at a spacetime point. Then, Eq.\ (\ref{ph_vel}) will admit two roots, given by
\begin{equation}
\label{eq_v_ph}
v_\varphi ^\pm = \frac{\beta \pm \sqrt{\beta^2 +4\alpha\gamma}}{2\alpha}.
\end{equation}
Complex roots are, in general, associated with the propagation of evanescent waves, which we do not analyze here. Therefore, we assume that the condition
\begin{equation}
    \beta^2 +4\alpha\gamma\geq 0
\end{equation}
holds for all possible directions in the rest space. Clearly, the latter leads to algebraic constraints which must be fulfilled by the constitutive tetrad at the corresponding point. An interesting feature of Eq.\ (\ref{ph_vel}) is that it is not, in general, invariant under spatial inversions. Indeed, a closer inspection of Eqs.\ (\ref{eq:18})-(\ref{eq:19}) reveals that
\begin{equation}
\hat{q}^{a}\mapsto-\hat{q}^{a},\quad\quad\quad v_\varphi ^\pm\mapsto -v_\varphi ^\mp.    
\end{equation}
However, in the particular case where magneto-electric cross terms are absent, the coefficient $\beta$ vanishes identically and one recovers the usual situation, where the ``effective light cone'' is invariant under reflection about the rest space. The particular case of linear isotropic dielectrics, for which $\mathfrak{A}^{a}{}_{b} = \varepsilon \, h^{a}{}_{b}$, $\mathfrak{B}^{a}=\mathfrak{C}_{a}=0$ and $\mathfrak{D}=1/\mu$, with $\varepsilon(x)$ and $\mu(x)$ as the dielectric functions, clearly illustrates this phenomena. Indeed, in this case, we recover the symmetric expression $v_\varphi^2 = (\mu\,\varepsilon)^{-1}$, which shows the mirror invariance.

It is worth mentioning that there is no room for birefringence in this theory due to the quadratic form of the dispersion relation encoded in Eq.\ (\ref{ph_vel}). However, for sufficiently high values (in modulus) of magneto-electric cross-terms, the theory predicts the existence of the phenomenon of \textit{one-way propagation}. Indeed, the latter will occur whenever the following inequalities hold
\begin{equation}
\label{bir_cond}
-\beta^2 < 4\alpha\gamma < 0,
\end{equation}
for a particular direction. We shall see next how this situation emerges in concrete examples.

\section{Polarization}
In the last section we have seen that the eigenvalue/eigenvector problem Eq.\ (\ref{eq:13}) will admit nontrivial solutions provided the phase velocity is a root of the quadratic equation Eq.\ (\ref{ph_vel}). Setting  the wave-polarization vector $e^{a}$ as a linear combination of the orthogonal space-like basis orthogonal to the observer $\{\hat q^{a},\hat{p}^{a}\}$, it follows that:
\begin{equation}
\label{eq:26}
    (v_\varphi^2\ \tilde{\mathfrak{A}}^{a}{}_{b} +v_\varphi\ A^{a}{}_{b} - \tilde{\mathfrak{D}}\,\hat p^{a}\hat p_{b})\,(a_{1}\hat q^{b} + a_{2} \hat p^{b}) = 0,
\end{equation}
where $a_{1}$ and $a_{2}$ are direction-dependent coefficients to be determined. Contracting the above equation respectively with $\hat{q}_{a}$ and $\hat p_{a}$ we find a set of two algebraic equations for $a_{1}$ and $a_{2}$. However, since the latter are constrained to satisfy Eq.\ (\ref{ph_vel}), we need to solve only one of them. In particular, choosing the equation corresponding to the Gauss law Eq.\ (\ref{eq:Gauss}) and assuming $v_{\varphi}\neq 0$, yields the relation
\begin{equation}
\label{polar_coef_relat}
a_{1}=-\left(\fracc{v_{\varphi}\, \tilde{\mathfrak{A}}_{ab}\hat q^{a}\hat p^{b}+\tilde{\mathfrak{B}}_{a}\hat q^{a}}{v_{\varphi}\,\tilde{\mathfrak{A}}_{ab}\hat q^{a}\hat q^{b}}\right)a_{2}.
\end{equation}
A remarkable consequence of the latter is that neither the electric permeability term $\tilde{\mathfrak{C}}_{a}$ nor the (inverse) permeability term $\tilde{\mathfrak{D}}$ contribute to the polarization in a $(2+1)$-dimensional theory. 

Furthermore, Eq.\ (\ref{polar_coef_relat}) reveals that the propagating  wave will be transversal, i.e.
\begin{equation}
\label{pol_lin_iso_diel}
e^{a}\perp\hat q^{a},
\end{equation}
whenever $a_{1}=0$ or, equivalently
\begin{equation}
v_{\varphi}\, \tilde{\mathfrak{A}}_{ab}\hat q^{a}\hat p^{b}+\tilde{\mathfrak{B}}_{a}\hat q^{a}=0.
\end{equation}
In particular, transversality is certainly guaranteed in the case of the above-mentioned linear isotropic dielectric, as expected. One might wonder whether the theory also admits longitudinal propagation. According to Eq.\ (\ref{polar_coef_relat}), this condition implies the relation $a_{2}=0$, which is the same as
\begin{equation}
\tilde{\mathfrak{A}}_{ab}\hat q^{a}\hat q^{b}=0.
\end{equation}
Up to now, it is not entirely clear to us if the fulfillment of the latter would somehow imply in an inconsistent propagation. 
   
\section{Special Cases}
In this section, we investigate three interesting nonlinear systems: i) purely magnetic material, ii) anisotropic/non-magnetic dielectric and, iii) second-order magneto-electric medium.

\subsection{Purely magnetic media}
\begin{figure}
\includegraphics[scale=0.3]{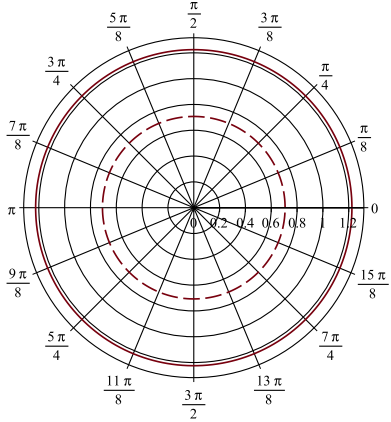}
\caption{Qualitative behavior of $v_{\varphi}$ for $B$ fixed. The solid line corresponds to $\bar\chi^{(2)}>0$ and the dashed line is for $\bar\chi^{(2)}<0$. For a given order in the expansion of $\mathfrak{D}(B)$, a positive susceptibility coefficient will give a bigger value for the phase speed in comparison to the negative one. (For the plots, we use $\bar\chi^{(1)}=0$, $\bar\chi^{(2)}=\pm1$ and $B=1/2$.)}
\label{pure_mag}
\end{figure}

We assume here that the (inverse) permeability tensor depends only on the magnetic field $\mathfrak{D}=\mathfrak{D}\,(B)$, while the other dielectric parameters are $\mathfrak{A}^{a}{}_{b} =\varepsilon_{0}\,h^{a}{}_{b}$ and $\mathfrak{B}^{a} = \mathfrak{C}_{a} = 0$. Thus, the non-vanishing components of the constitutive tetrad, obtained from Eqs.\ (\ref{tildeA})-(\ref{tildeD}), reduce to \begin{equation}
\tilde{\mathfrak{A}}^{a}{}_{b}=\varepsilon_{0}\,h^{a}{}_{b}\quad \mbox{and} \quad \tilde{\mathfrak{D}}=\mathfrak{D}+\frac{\partial\mathfrak{D}}{\partial B}\,B.
\end{equation}
Under these conditions, the expression for the phase velocity, given by Eq.\ (\ref{eq_v_ph}), becomes
\begin{equation}
\label{ph_vel_mag}
v_{\varphi}=\pm\sqrt{\frac{\gamma}{\alpha}}=\pm\sqrt{\frac{(\mathfrak{D}\,B)'}{\varepsilon_{0}}},
\end{equation}
    where (\,$'$\,) means here derivative with respect to B. This equation does not involve the spatial direction of the wave vector $\hat q^{a}$, indicating that $v_{\varphi}$ is isotropic, for any given $B$. Furthermore, if $\mathfrak{D}(B)$ admits a Taylor expansion, for each order of the power series, the corresponding susceptibility coefficient will increase or decrease the phase velocity, depending if it is positive or negative. For a better illustration of this phenomena, we refer the reader to  Fig.\ (\ref{pure_mag}), where we have depicted $v_{\varphi}(B)$ taking into account corrections up to second-order susceptibility coefficients \cite{born_wolf}, i.e.
\begin{equation}
\mathfrak{D}(B)=\mu_0^{-1}\left(1+\bar\chi^{(1)}+\bar\chi^{(2)}B\right).
\end{equation}

The wave-polarization vector $e^{a}$ can be directly obtained from Eq.\ (\ref{polar_coef_relat}), which coincides with the wave-polarization vector of linear isotropic dielectrics, given by Eq.\ (\ref{pol_lin_iso_diel}). Recall that it is orthogonal to the spatial direction of the wave vector.

\subsection{Anisotropic and non-magnetic dielectric media}

Now, we assume that the permittivity tensor depends only on the norm of the electric field, with independent behavior along each spatial direction, namely,
\begin{equation}
\mathfrak{A}^{a}{}_{b}=\mbox{diag}\,[0,\varepsilon_x(||E||),\varepsilon_y(||E||)].
\end{equation}
The other dielectric and magneto-electric parameters are chosen as $\mathfrak{D} = \mu_0^{-1}$ and $\mathfrak{B}^{a}=\mathfrak{C}^{a}=0$. Thus, the only non-vanishing components of the constitutive tetrad are 
\begin{equation}
\tilde{\mathfrak{A}}^{a}{}_{b}=\mathfrak{A}^{a}{}_{b}+\frac{\partial\mathfrak{A}^{a}{}_{c}}{\partial ||E||}\, \frac{E^{c} E_{b}}{||E||}\quad \mbox{and} \quad \tilde{\mathfrak{D}}=\mu_0^{-1}.
\end{equation}

\begin{figure}
\includegraphics[scale=0.3]{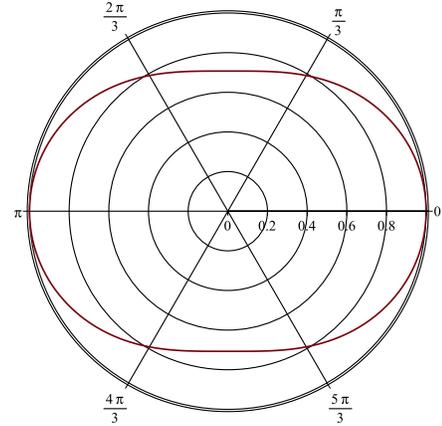}
\caption{Plot of $v_{\varphi}$ as function of $\phi$. For the illustrative choice $\varepsilon_y=\varepsilon_0$, $\varepsilon_x=\varepsilon_0||E||$ with $||E||=1.0$, one sees that the phase speed decreases from its maximum value along the $x$-axis to its minimum along the $y$-axis.}
\label{anis-dielectric}
\end{figure}

For the sake of simplicity, we can set the electric field along the $x$-direction and decompose the unit spatial wave vector as $\hat q_{a}=(0,\cos\phi,\sin\phi)$, where $\phi$ is measured with respect to the $x$-axis. Under these assumptions, the expression (\ref{eq_v_ph}) for the phase speed reduces to
\begin{equation}
\label{eq_v_ph_anis}
v_\varphi = \pm\sqrt{\frac{\gamma}{\alpha}} = \pm\sqrt{\frac{\cos^2\phi}{\mu_0\,\varepsilon_y} + \frac{\sin^2\phi}{\mu_0(||E||\varepsilon_x)'}},
\end{equation}
where prime (\,$'$\,) means here derivative with respect to $||E||$. Note that we have only one possibility for the phase speed, in modulo, in addition to the fact the light ray is always \textit{extraordinary} \cite{Teodoro2004}, that is, it depends on the wave vector direction. For $||E||$ fixed, Eq.\ (\ref{eq_v_ph_anis}) determines the surface normal of the waves (see Fig.\ \ref{anis-dielectric}). 

\begin{figure}
\includegraphics[scale=0.21]{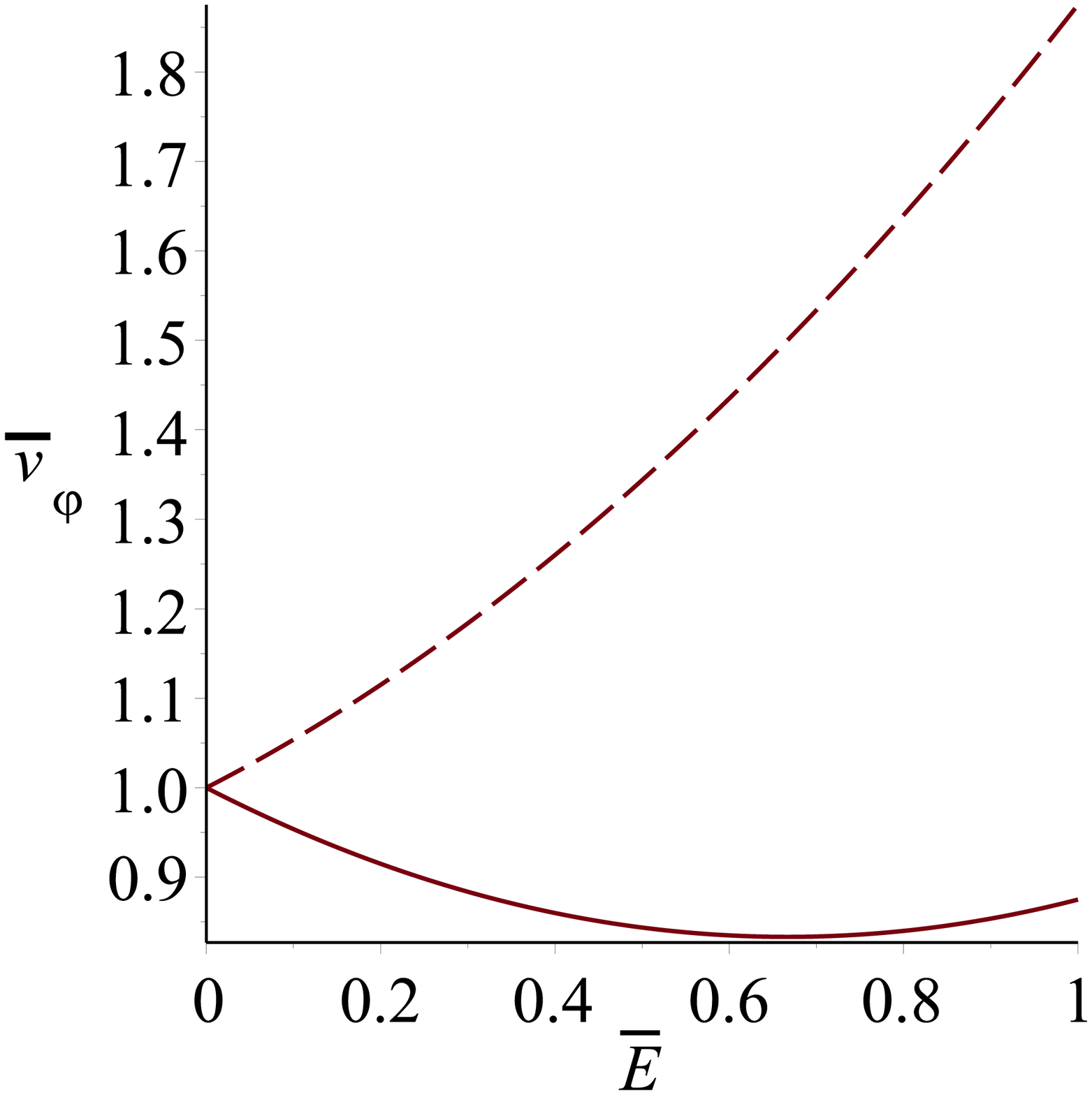}
\includegraphics[scale=0.21]{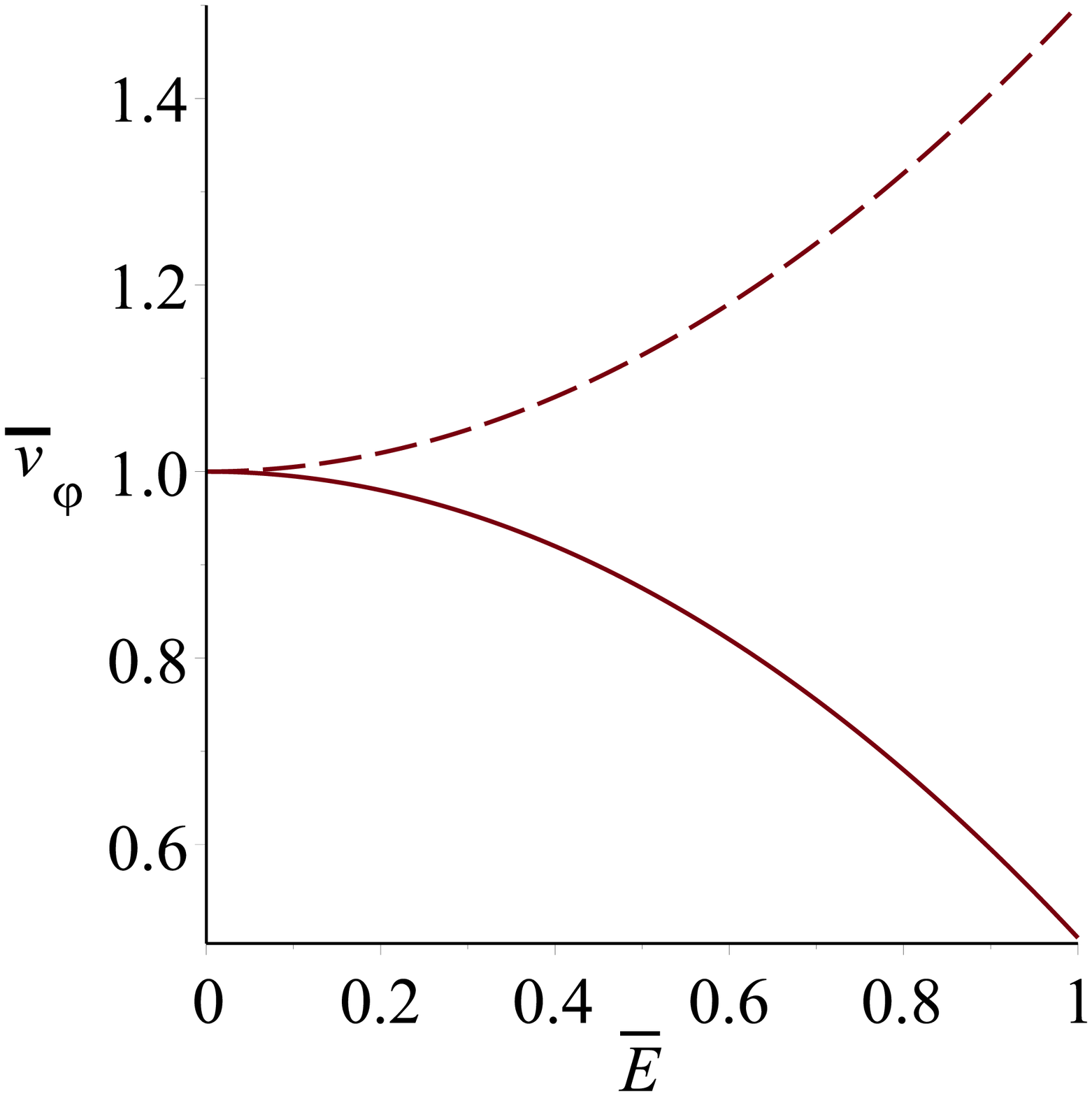}
\caption{Plots of $v_{\varphi}(||E||)$. On the left-hand side, the solid line corresponds to a positive $\chi_y^{(2)}$, with the phase speed having a minimum value as $||E||$ increases. The dashed line represents an always increasing phase speed for $\chi_y^{(2)}<0$. On the right-hand side, the solid line corresponds to a positive $\chi_y^{(3)}$, yielding a decreasing phase speed, while the dashed line indicates a monotonically increasing $v_{\varphi}$ for $\chi_y^{(3)}<0$. For convenience, we have defined $\bar v_{\varphi}=\sqrt{\varepsilon_0(1+\chi_y)}v_{\varphi}$ and $\tilde E=\frac{\chi_y^{(2,3)}}{\varepsilon_0(1+\chi_y)}||E||$, correspondingly.}
\label{chi_anis}
\end{figure}

For a better illustration of such effect, we shall study two particular cases, assuming that the permittivity functions $\varepsilon_x$ and $\varepsilon_y$ admit a power-law expansion in $E$. Thus,
\begin{eqnarray}
\varepsilon_x=\varepsilon_0(1 + \chi_{x}^{(1)} + \chi_{x}^{(2)}||E| + \chi_{x}^{(3)}||E||^2) + {\mathcal O}(3),\\[1ex]
\varepsilon_y=\varepsilon_0(1 + \chi_{y}^{(1)} + \chi_{y}^{(2)}||E|| + \chi_{y}^{(3)}||E||^2) + {\mathcal O}(3),
\end{eqnarray}
where $\chi_{x,y}^{(n)}$ are the $n$-th order susceptibility coefficients. In principle, these coefficients can have both signs, but the change in the phase speed is qualitatively the same at certain order. The linear electro-optic effect (``Pockel-like''), due to $\chi^{(2)}$-corrections, makes the phase speed increase\red{s} ($\chi^{(2)}<0$) or decrease\red{s} ($\chi^{(2)}>0$) for small values of $||E||$. But, for large $||E||$, both signs lead to a quadratic increasing in $v_{\varphi}$ (see Fig.\ \ref{chi_anis}). The quadratic electro-optic effect (``Kerr-like'') caused by the presence of $\chi^{(3)}$ does not allow such a change in the profile of $v_{\varphi}(||E||)$, which always increases ($\chi^{(3)}<0$) or decreases ($\chi^{(3)}>0$) quadratically as $||E||$ grows (see Fig.\ \ref{chi_anis}). 

From Eq.(\ref{polar_coef_relat}), it is straightforward to determine the form of the polarization modes allowed for this case. Since the magneto-electric coefficients are absent, the wave-polarization vector $e^{a}$ is simply given by
\begin{equation}
    e^{a} = \left(0,\frac{\sin\phi}{(E\,\varepsilon_{x})'},-\frac{\cos\phi}{\varepsilon_{y}}\right)
\end{equation}

\subsection{Second order magneto-electric media}\label{mag-elec_media_sec}
The magneto-electric media we shall study now is the 2D adaptation of the 3D medium analyzed in Ref.\ \cite{DeLorenci2022}, where the electric displacement and the magnetic induction are obtained from first principles through a Taylor expansion of the Helmholtz free energy in terms of the electric and magnetic fields. The same is done here (see details in Appendix \ref{excit_from_helm}), leading to the following polarization vector and magnetization scalar
\begin{eqnarray}
&&P_{a}=\varepsilon_0\chi_{ab}^{(1)} E^{b} + \frac{1}{2}\chi_{a}^{(2)}H^2,\\[1ex]
&&M=\tilde\chi^{(1)} H + \mu_0^{-1}\chi_{a}^{(2)}E^{a}H,
\end{eqnarray}
where the susceptibility coefficients $\chi_{ab}^{(1)}$ and $\tilde\chi^{(1)}$ are first order contributions to the polarization and magnetization, respectively, while the susceptibility vector $\chi_{a}^{(2)}$ is a second order correction for both induced fields. Using the relations between $P_{a}$ and $M$ with the field excitations, namely, $D_{a}=\varepsilon_0 E_{a}+P_{a}$ and $B=\mu_0(H+M)$, one can directly read the constitutive tensor, vectors and scalar defined in Eqs.\ (\ref{const_relat1}) and (\ref{const_relat2}), as follows
\begin{eqnarray}
&&\mathfrak{A}_{ab}=\varepsilon_0[h_{ab} +\chi_{ab}^{(1)}],\quad \mathfrak{B}_{a}=\fracc{1}{2}B\,\mathfrak{D}^2\,\chi_{a}^{(2)},\\[1ex]
&&\mathfrak{C}^{a}=0\quad\mbox{and}\quad\mathfrak{D} =[\mu_0(1+\tilde\chi^{(1)}) + \chi_{a}^{(2)}E^{a}]^{-1}. \end{eqnarray}
For this case, the elements of the constitutive tetrad are
\begin{eqnarray}
&&\tilde{\mathfrak{A}}_{ab}=\mathfrak{A}_{ab}-B^2\mathfrak{D}{}^3 \chi_{a}^{(2)}\chi_{b}^{(2)}, \quad \tilde{\mathfrak{B}}^{a}=2\mathfrak{B}^{a},\\[1ex]
&&\tilde{\mathfrak{C}}^{a}=-2\mathfrak{B}^{a}\quad \mbox{and} \quad \tilde{\mathfrak{D}} = \mathfrak{D}.
\end{eqnarray}

We shall assume again the electric field along the $x$-axis, decompose the unit spatial wave-vector in polar coordinates and write $\chi_{a}^{(2)}=(0,\chi_1,\chi_2)$. Thus, the phase speed equation (\ref{eq_v_ph}) becomes
\begin{equation}
\sqrt{\mu\varepsilon}\,v_{\varphi}^{\pm}=\frac{(\tilde E+1)\left[\tilde B(r\cos\phi-\sin\phi)\pm\sqrt{\Delta}\right]}{(\tilde E+1)^3-(1+r^2)\tilde B^2}
\end{equation}
where $\tilde E=\chi_1\,E/\mu$, $\tilde B=B\chi_1/\sqrt{\varepsilon\mu^3}$, $r=\chi_2/\chi_1$ and $\Delta=(\tilde E+1)^3 -  \tilde B^2(\cos\phi + r\sin\phi)^2$, with $\varepsilon=\varepsilon_0[1+\chi^{(1)}]$ and $\mu=\mu_0[1+\tilde\chi^{(1)}]$. 

\begin{figure}
\includegraphics[scale=0.3]{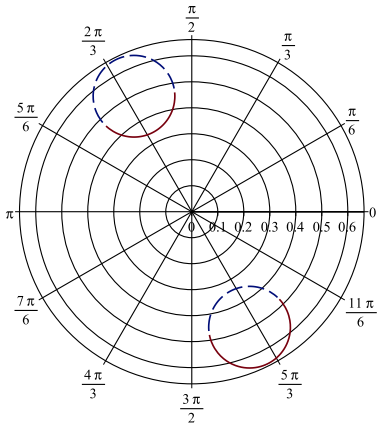}
\caption{Phase speeds $v_{\varphi}^{+}$ (solid line) and $v_{\varphi}^{-}$ (dashed line), in modulo, as function of $\phi$. It is possible to see the complementary windows of one-way propagation and opacity. For this, we choose $\tilde B = 1.0$, $\tilde E = -0.5$ and $r=0.5$.}
\label{vel_bir_yes}
\end{figure}

\begin{figure}
\includegraphics[scale=0.3]{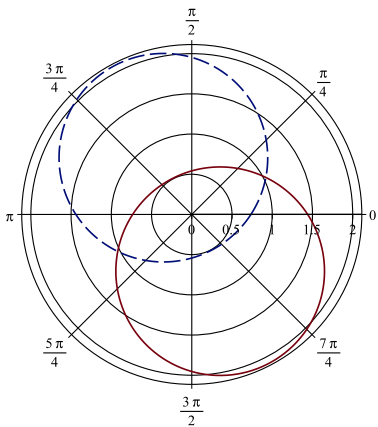}
\caption{Phase speeds $v_{\varphi}^{+}$ (solid line) and $|v_{\varphi}^{-}|$ (dashed line), in modulo, as function of $\phi$. For the choice $\tilde B = 1.0$, $\tilde E = 0.5$ and $r=0.5$, one of the conditions for the one-directional propagation is not satisfied and then we have $v_{\varphi}^{+}$ and $v_{\varphi}^{-}$ for all directions.}
\label{vel_bir_no}
\end{figure}

From the equation above, one can read the condition for the one-way propagation as
\begin{equation}
\Delta>0\quad \mbox{and}\quad (1+\tilde{E})^{3}< \tilde{B}^{2}\,(1+r^{2}).
\end{equation}
Note that the first inequality has an angular dependence on the direction $\phi$ of the wave vector, while the second inequality involves only the field strengths. The former provides a window for the one-way propagation from the roots of the quadratic equation
\begin{equation}
    [r^{2}\,\tilde{B}^{2}-(1+\tilde{E})^{3}]\,\tan^2\phi + 2\,r\,\tilde{B}^{2}\,\tan\phi+\tilde{B}^{2}-(1+\tilde{E})^{3}= 0,
\end{equation} 
which are
\begin{equation}
    \tan\phi^{\pm}_c= \frac{-r\,\tilde{B}^{2}\pm\sqrt{[\tilde{B}^{2}\,(r^{2}+1)-(1+\tilde{E})^{3}]\,(1+\tilde{E})^{3}}}{[r^{2}\,\tilde{B}^{2}-(1+\tilde{E})^{3}]}.
\end{equation}
Therefore, the phenomenon of the one-way propagation exists for $\tan\phi^{-}_c<\tan\phi<\tan\phi^{+}_c$, which is depicted in Fig.\ \ref{vel_bir_yes}. Note that out of this window, the medium is opaque. In Fig.\ \ref{vel_bir_no}, we show that it is possible to have light propagating for all spatial directions by changing only the ratio $r$, with $E$ and $B$ fixed, but the propagation is anisotropic. From both figures, it is evident the symmetry $v_{\varphi}^{\pm}$ into $-v_{\varphi}^{\mp}$ under the transformation $\phi\rightarrow \phi+\pi$, as mentioned before, recalling that in polar plots, the radius represents the magnitude of the quantity, without taking into account its sign. 

\begin{figure}
\includegraphics[scale=0.9]{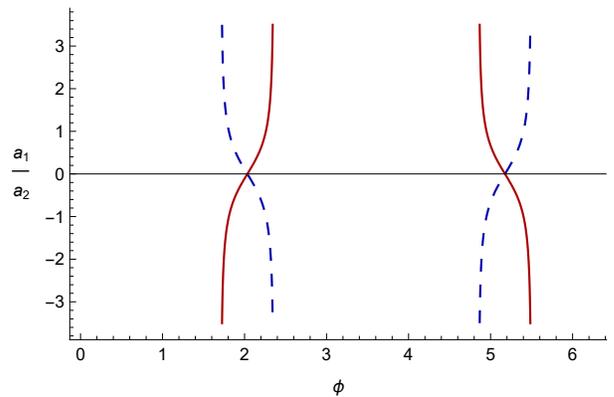}
\caption{The ratio $a_1/a_2$ as function of $\phi$. This is an alternative way to see windows of one-way propagation and opacity. The transformation $\phi\rightarrow \phi+\pi$ evinces now the map between the wave-polarization vectors of  $v_{\varphi}^{+}$ and $v_{\varphi}^{-}$. We choose here $\tilde B = 1.0$, $\tilde E = -0.5$ and $r=0.5$.}
\label{polar_fig}
\end{figure}

Concerning the wave-polarization vector in this case, Eq.\ (\ref{polar_coef_relat}) can be written down as
\begin{equation}
a_{1} = -\frac{(1+\tilde{E})^{3}}{\varepsilon\,\Delta}\left(\frac{\tilde{B}_{a}\hat{q}^{a}}{v_{\varphi}{}^{\pm}}-\tilde{B}_{a}\hat{q}^{a}\tilde{B}_{b}\hat{p}^{b}\right)a_{2}.
\end{equation}
Since the explicit form of this equation is a cumbersome, we depict the behavior of the ratio $a_1/a_2$ in Fig.\ \ref{polar_fig}. There, one can see once again, the spatial symmetry in flipping the direction of $\hat q^{a}$ and also the distinction with respect to the phenomenon of birefringence: when $\phi\rightarrow \phi+\pi$ the wave-polarization vector of $v_{\varphi}^{+}$ is mapped onto the wave-polarization vector of $v_{\varphi}^{-}$ and vice versa, rendering impossible the configuration of two independent phase velocities with independent wave-polarization vectors for a given direction.

\section{Concluding remarks}
We presented the geometrical aspects of the light propagation inside nonlinear 2D material media subject to external electromagnetic fields. Starting with the Faraday tensor as a fundamental 2-form, we constructed the field equations in a $2+1$-dimensional spacetime in terms of the divergence of the electromagnetic field and the corresponding Bianchi identity.

The geometric optic limit inside a nonlinear 2D material is achieved by applying the Hadamard theorem to the characteristic surface, obtaining the characteristic equation, the dispersion relation and, finally, the expression for the phase speed. We also derived the general form of the wave-polarization vector. Thus, we examined three distinct cases of interest, calculating the phase speed and the wave-polarization vector for each of them. Despite of the dimension reduction, we proved the existence of one-way propagation and controlled opacity in such media, particularly for the magneto-electric ones. 

Contrary to the case of linear electromagnetic theories in vacuum \cite{McDonald2019,Wheeler1997,DBoito2020}, it is possible to see, from the beginning, the non-equivalence between the nonlinear electrodynamics in $3+1$-dimension and the $2+1$-dimensional formulation developed here. In particular, the constitutive relations (\ref{const_relat1}) and (\ref{const_relat2}) must involve objects with different tensor ranks, consequently, the number of degrees of freedom is distinct from the $3+1$ case. This might be an important point to distinguish both formulation and possibly favor one over the other in the description of 2D nonlinear materials.

In virtue of the observer-independent formulation developed in our previous work, the effective optical metric of a nonlinear 2D material can be directly obtained from equation (37) of Ref.\ \cite{bit_bran_goul_22}, by replacing the constitutive tetrad basis of a linear medium by the one associated with a nonlinear medium, given by Eqs.\ (\ref{tildeA})-(\ref{tildeD}). In this vein, the next step will be the investigation of analogue models of gravity in this context. 

\acknowledgments
We would like to thank the participants of the \textit{GIFT Seminar}, specially Vitorio De Lorenci for many discussions that helped us to improve the manuscript. EOSB thanks CNPq for the financial support (Grant number 134395/2021-2).

\appendix

\section{Polarization and magnetization vectors of a nonlinear medium at thermodynamic equilibrium}\label{excit_from_helm}

Following the standard thermodynamic approach, the free-energy  density $\mathcal{F}$ of a 2D material in the presence of electromagnetic fields can be written as function of the electric field $E^{\mu}$ and the magnetic excitation $H$, at a fixed temperature $T$; that is
\begin{equation}
    \mathcal{F}=\mathcal{F}(E^{\mu},H;T).
\end{equation}
Needless to say, this quantity is not covariant, in sense that its definition is frame-dependent, since it mixes field strengths with field excitations.

As long the fields $E^{\mu}$ and $H$ are bounded, for practical reason, we can expand $\mathcal{F}$ in terms of those fields (up to third order) as 
\begin{equation}
\begin{array}{lcl}
\mathcal{F}&=&\mathcal{F}_0 - P^{(s)}_{a}E^{a} - M^{(s)}H - \fracc{1}{2}\varepsilon_0\chi^{(1)}_{ab}E^{a}E^{b}\\[2ex]
&& - \fracc{1}{2}\mu_0\bar\chi^{(1)} H^2 - \alpha^{(1)}_{a}E^{a}H - \fracc{1}{3}\varepsilon_0\chi^{(2)}_{abc}E^{a}E^{b}E^{c} \\[2ex]
&& - \fracc{1}{3}\mu_0\bar\chi^{(2)}H^3  - \fracc{1}{2}\alpha^{(2)}_{ab}E^{a}E^{b}H - \fracc{1}{2}\beta^{(2)}_{a}E^{a}H^2,
\end{array}
\end{equation}
where $P^{(s)}$ and $M^{(s)}$ stand for spontaneous polarization and magnetization effects, respectively. All the other tensor coefficients are the susceptibilities of the material, with the subscript number in the  parentheses indicating the corresponding order in the power expansion of $\mathcal{F}$.

Hence, the polarization vector can be calculated and it is written down as
\begin{equation}
\begin{array}{lcl}
P^{a}\doteq-\fracc{\partial\mathcal{F}}{\partial E^{a}}&=& P^{(s)}_{a} +\varepsilon_0\chi^{(1)}_{ab}E^{b} +\alpha^{(1)}_{a}H\\[2ex]
&&+ \varepsilon_0\chi^{(2)}_{abc}E^{b}E^{c} + \alpha^{(2)}_{ab}E^{b}H + \fracc{1}{2}\beta^{(2)}_{a}H^2.
\end{array}
\end{equation}
Analogously, the magnetization is given by
\begin{equation}
\begin{array}{lcl}
\mu_0M\doteq-\fracc{\partial\mathcal{F}}{\partial H}&=&
M^{(s)} + \mu_0\bar\chi^{(1)} H + \alpha^{(1)}_{a}E^{a} + \mu_0\bar\chi^{(2)} H^2 \\ [2ex] 
&& + \fracc{1}{2}\alpha^{(2)}_{ab}E^{a}E^{b} + \beta^{(2)}_{a}E^{a}H.
\end{array}
\end{equation}

Thus, the electric displacement $D_{a}=\varepsilon_0 E_{a}+P_{a}$ is obtained through
\begin{equation}
\label{D_fnc_E}
\begin{array}{lcl}
D_{a}&=& \varepsilon_0\left(\delta_{ab} + \chi^{(1)}_{ab} + \chi^{(2)}_{abc}E^{c} + \frac{1}{\varepsilon_0}\alpha^{(2)}_{ab}H\right)E^{b}\\[2ex]
&&+\left(\alpha^{(1)}_{a} + \fracc{1}{2}\beta^{(2)}_{a}H\right)H,
\end{array}
\end{equation}
while the magnetic strength $B=\mu_0(H+M)$ can be written as
\begin{equation}
\label{B_fnc_H}
\begin{array}{lcl}
B&=& \mu_0\left(1 + \bar\chi^{(1)} + \bar\chi^{(2)}H + \frac{1}{\mu_0}\beta^{(2)}_{a}E^{a}\right)H\\[2ex] &&\left(\alpha^{(1)}_{a}+\fracc{1}{2}\alpha^{(2)}_{ab}E^{b}\right)E^{a}.
\end{array}
\end{equation}
From this, one sees that Eq.\ (\ref{B_fnc_H}) is a polynomial equation in $H$, which should be solved in order to find the constitutive relation $H=H(E^{a},B)$. In possession of it, we can substitute $H=H(E^{a},B)$ in Eq.\ (\ref{D_fnc_E}) to find the other constitutive relation $D^{a}=D^{a}(E^{b},B)$. Only after such manipulation, the dielectric and magneto-electric coefficients can be determined. For the sake of illustration, we study the case $\bar\chi^{(2)}=0$ in Sec.\ \ref{mag-elec_media_sec}, such that $B$ and $H$ are linearly related.


\bibliography{refs}

\end{document}